%% file: main.tex
\documentclass[conference]{IEEEtran}
\IEEEoverridecommandlockouts
\usepackage[letterpaper,left=18mm,right=18mm,top=16mm,bottom=16mm]{geometry} 
\usepackage{amsmath,amssymb,amsfonts}
\usepackage{algorithmic}
\usepackage[pdftex]{graphicx}
\usepackage{textcomp}
\usepackage{xcolor}
\usepackage{tabularx}
\usepackage{authblk}
\usepackage{svg}
\usepackage{listings}
\usepackage{enumitem}
\usepackage[font=footnotesize]{caption}
\usepackage{footnote}
\usepackage{array}
\newcolumntype{L}[1]{>{\raggedright\arraybackslash}m{#1}}
\newcolumntype{C}[1]{>{\centering\arraybackslash}m{#1}}
\newcolumntype{R}[1]{>{\raggedleft\arraybackslash}m{#1}}
\usepackage{varwidth}

\lstdefinestyle{lstStyleBase}{
	,basicstyle=\scriptsize\ttfamily 
	,floatplacement=tbp    
	,aboveskip=\smallskipamount 
	,belowskip=\smallskipamount 
	,lineskip=0pt          
	,boxpos=c              
	,showlines=false       
	,extendedchars=true   
	,upquote=true         
	,tabsize=2,           
	,showtabs=false       
	,showspaces=false     
	,showstringspaces=false 
	,numbers=none         
	,stepnumber=2         
	,numberfirstline=false 
	,numberstyle=\tiny\color{blue}    
	,numbersep=5pt        
	,numberblanklines=true %
	,numberbychapter=true %
	,captionpos=b         
	,abovecaptionskip=\smallskipamount 
	,belowcaptionskip=\smallskipamount 
	,linewidth=\linewidth 
	,xleftmargin=0pt      
	,xrightmargin=0pt     %
	,resetmargins=false   
	,breaklines=true      
	,breakatwhitespace=false 
	,breakindent=0pt     
	,breakautoindent=true 
	,columns=flexible     %
	,keepspaces=true      %
	,frame=lines         
	,framesep=3pt 
	,rulesep=2pt          
	,framerule=0.4pt      
	,language=C
}

\newcommand{\dirquote}[1]{``#1''}

\usepackage[linesnumbered,ruled,vlined]{algorithm2e}

\SetCommentSty{mycommfont}
\SetKwInput{KwInput}{Input}                
\SetKwInput{KwOutput}{Output}              

\def\BibTeX{{\rm B\kern-.05em{\sc i\kern-.025em b}\kern-.08em
    T\kern-.1667em\lower.7ex\hbox{E}\kern-.125emX}}

\begin{document}

\title{Agile Autotuning of a Transprecision Tensor Accelerator Overlay for TVM Compiler Stack} 

%

\author[1]{Dionysios Diamantopoulos}
\author[1]{Burkhard Ringlein}
\author[1]{Mitra Purandare}
\author[2]{Gagandeep Singh}
\author[1]{Christoph Hagleitner}
\affil[1]{IBM Research Europe, S\"aumerstrasse 4, 8803 R\"uschlikon, Switzerland} 
\affil[2]{Eindhoven University of Technology, Netherlands}

\maketitle
\begin{abstract}
Specialized accelerators for tensor-operations, such as blocked-matrix operations and multi-dimensional convolutions, have been emerged as powerful architecture choices for high-performance Deep-Learning computing. The rapid development of frameworks, models, and precision options challenges the adaptability of such tensor-accelerators since the adaptation to new requirements incurs significant engineering costs. Programmable tensor accelerators offer a promising alternative by allowing reconfiguration of a virtual architecture that overlays on top of the physical FPGA configurable fabric. We propose an overlay ($\tau$-VTA) and an optimization method guided by agile-inspired auto-tuning techniques. We achieve higher performance and faster convergence than state-of-art.

\end{abstract}


\input{intro}
\input{method}

\input{optimize}

\input{expt}

\input{related}

\input{conclusion}

\bibliographystyle{plain}
\bibliography{main}

\end{document}

%% file: intro.tex
\section{Introduction}
Deep Learning (DL), a powerful set of techniques for learning in neural networks, has achieved unprecedented accuracy in numerous aspects of the digital transformation of our society. This fascinating biologically-inspired programming paradigm, which enables a computer to learn from observational data, attributes its success to a large volume of trained parameters, which, however, can contain a lot of redundant information~\cite{DBLP:journals/corr/HanMD15}. Prior art has maintained remarkable levels of accuracy, by applying pruning techniques~\cite{10.1007/978-3-319-70096-0_41}\cite{DBLP:journals/corr/HanMD15}\cite{8192500}, or by using sparsification~\cite{7298681}, or both~\cite{Han:2015:LBW:2969239.2969366}\cite{Iandola2016}. A highly effective technique exploiting redundant information is moving from floating-point arithmetic to low-precision integer arithmetic~\cite{Blott2018}. 
Recently, transprecision computing was proposed as a paradigm shift in precision selection, and suggests the adaptability of precision according to an application's requirements~\cite{8342176}, as opposed to the conservative static selection of low precision processing.

DL systems rely on hardware (HW) accelerators and manually optimized, high-performance libraries to increase computational efficiency, i.e., achieving the maximum throughput while consuming the smallest possible amount of resources and energy~\cite{Li2017,Pouyanfar:2018:SDL:3271482.3234150}. While GPUs are dominating the training and inference computations at scale, FPGAs can achieve more than 10$\times$ better speed and energy efficiency than state-of-the-art GPUs \cite{10.1145/3289185,Mittal2018}. In addition, the inherent programmability of FPGAs at bit-level makes them ideal accelerators for ultra low-precision inference and training~\cite{9026948, LIANG20181072}. 

To optimize a neural network, programmers must choose from many implementations that are logically equivalent but differ dramatically in performance due to differences in threading, memory reuse, pipelining and other HW factors. Supporting diverse HW back-ends therefore incurs significant engineering cost. Depending on whether it is an FPGA, a GPU or an ASIC, the accelerator's adaptability to different models heavily relies on the software-supported libraries that bridge the gap between the programming language semantics and the HW supported intrinsics. A recent study proposed the use of a statistical cost model that predicts program run time by using a given low-level program~\cite{10.5555/3327144.3327258}. The cost model guides the exploration of the space of possible programs. 

{\setlength{\belowcaptionskip}{-4ex}
\begin{figure}[t]
\includegraphics[width=0.5\textwidth]{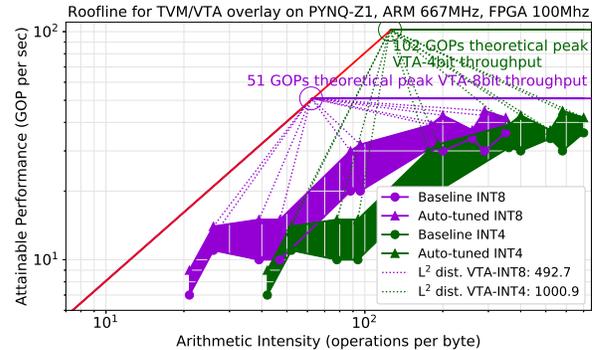}
\caption{Auto-tuning a model for an overlay FPGA back-end does not necessarily lead to higher performance, when the precision is decreased (e.g. INT8 to INT4). Investigating the reason and proposing path-forward motivates this work.}
\label{fig1}
\end{figure}
}

While auto-tuning has been proved to be effective in automatic optimization, it typically assumes a fixed HW that offers tunable knobs in the software, like tiling, loop reordering, etc. However, in the case of FPGAs, the HW design space also provides both tunable micro-architecture choices, e.g. pipelining, and tunable implementation choices, e.g. device frequency.
Such HW tunable choices can lead to sub-optimal auto-tuning when only a subset of them is considered for bitstream generation, particularly in low-precision DL for FPGAs~\cite{Blott2018}. 

To illustrate this problem, we examine the auto-tuning of an FPGA overlay developed by the community-driven TVM DL compiler stack~\cite{tvm}, specifically VTA~\cite{8764458}. This overlay employs a GEMM accelerator as a computation engine for convolution operations. We configured a GEMM of 8 bits (W8A8) and 4 bits (W4A4), for weights and activations, respectively. Figure~\ref{fig1} depicts the roofline model \cite{Williams:2009:RIV:1498765.1498785} of VTA, overlaid on a PYNQ-Z1 device~\cite{pynq}. The plot shows the throughput achieved on different convolution layers of the ResNet-18 inference benchmark. Each layer has a different arithmetic intensity, i.e. compute to data movement ratio. In the left half of the plot, convolution layers are bandwidth-limited, whereas on the right half, they are compute-limited. The operation points on the roofline, depicted as circles in Figure~\ref{fig1}, are ``optimal'' since in those points  neither performance nor communication is under-utilized. The goal behind designing HW architectures and compiler stacks is to bring each workload as close as possible to the roofline of the target HW and ideally to these ``optimal'' points. 

As an experiment, we auto-tuned the VTA, using TVM's auto-tuning flow, configured with a GEMM of W8A8 and W4A4. When VTA is configured with a 4-bit GEMM intrinsic, the theoretical performance is doubled to 102GOPs. In addition, the arithmetic intensity is doubled since half the number of bytes have to be fetched from the main memory. Please note that this is a strong advantage of overlay architectures on FPGAs~\cite{Blott2018}. 
However, the case of W4A4 delivers a measured performance identical to that of VTA W8A8, which wastes 100\% of the theoretical possible speedup. In addition the Euclidean distance ($L^{2}$) of all auto-tuned convolutions is higher for W8A8 than W4A4 from the perspective of the ``optimal''  operation points for VTA of 8 bit and 4 bits. This shortcoming is attributed to the current TVM compiler stack that relies on the user for the selection of the HW parameters of the VTA overlay for different precision. In addition, the VTA auto-tuning, using TVM's current auto-tuning flow, focuses only on software-related optimization options, assuming a fixed VTA design on the FPGA. While this assumption enables the interoperability of TVM's auto-tuning flow to silicon-proven devices, such as GPUs, DSPs and custom ASICs, it neglects some important features of reconfigurable HW.

In this paper, we explore the following question: \textbf{Can we automatically guide the auto-tuning of a tensor accelerator overlay, for different precision settings, by leveraging knowledge from hardware design experience?} Our affirmative answer is based on a framework that makes the following contributions:

\begin{itemize}[label=$\diamond$, wide]
\item \textit{Engineering aspect:} We adopt the concept of agile development to the community-driven TVM github repository, by bringing-up a pipeline of engineering tasks that extends the autotuning process. When this step is finished, the best explored models can be uploaded to benefit the community.
\item \textit{Scientific aspect:} Instead of eliminating the overlay hardware design space with pruning techniques (e.g. successive halving, used in~\cite{8764458}), we propose a technique that builds a prediction model that quantifies the impact of a hardware design choice (feature) towards an optimization goal, e.g. increasing performance. 
By employing a classifier with an arbitrary differentiable loss function, we show that the features with the highest impact differ for different precisions. We further propose the use of the most important features in order to generate an overlay and then continue with auto-tuning.
\end{itemize}

In Section~\ref{agile} we present the design-automation engineering contribution, i.e. the agile auto-tuning methodology. To demonstrate the impact of this new integral development concept, we propose the $\tau$-VTA optimization in Section~\ref{tvta}.
Experimental results are presented in Section~\ref{results} and Section~\ref{conclusion} concludes the paper.

%% file: method.tex
\section{Agile Autotuning Methodology}\label{agile}

\subsection{Integration to TVM}

To handle the expanding ecosystem of DL Frameworks on the one side and specialized DL HW on the other, a group from the University of Washington proposed \textit{TVM}, a full open source compiler-stack \cite{10.5555/3291168.3291211,8764458,Roesch2018} that aims to \dirquote{close the gap between the productivity-focused deep learning frameworks, and the performance- or efficiency-oriented hardware backends}~\cite{tvm}. TVM is built using multiple Intermediate Representation (IR) languages and therefore offers multiple layers for optimizations, as seen on the left-hand side of Figure~\ref{fig2}.
It first has multiple modules to import state-of-the-art DL Frameworks (like pytorch, Tensorflow, Keras) to the RelayIR~\cite{Roesch2018}. Afterwards, it is able to perform multiple optimizations on the Abstract Syntax Tree (AST) generated out of RelayIR and lower the program to the TVM IR. The TVM IR can then be interpreted by a runtime or another compiler to finally execute the DL task on the target HW~\cite{8764458}.

{\setlength{\belowcaptionskip}{-2ex}
\begin{figure}[t]
\includegraphics[width=0.5\textwidth]{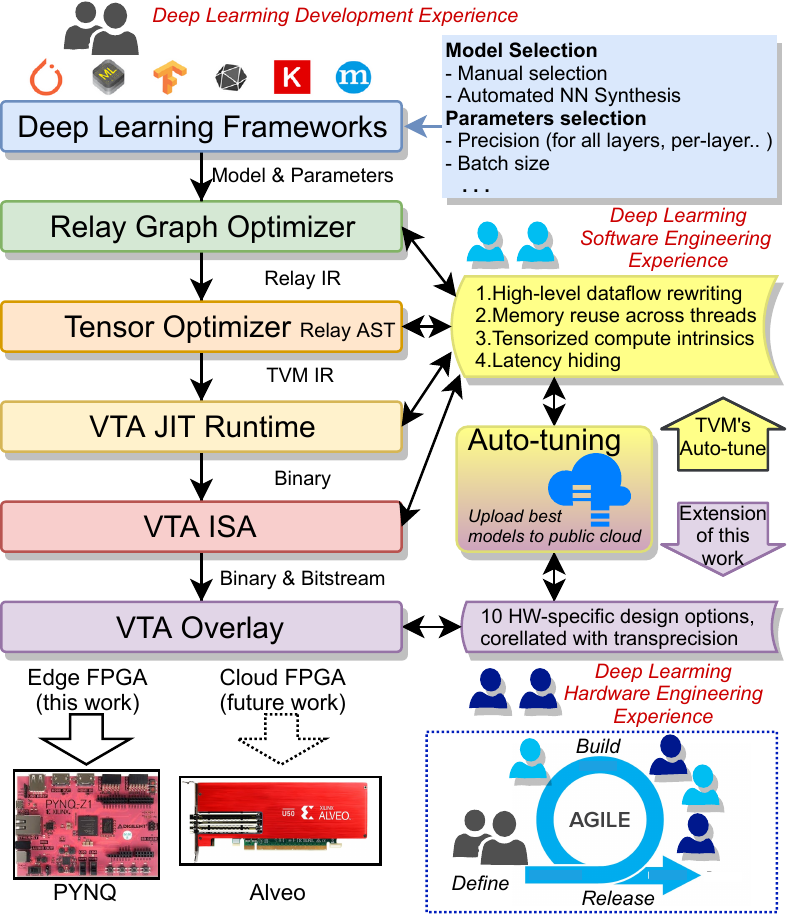}
\caption{Overview of the proposed approach: the auto-tuning step of the TVM toolflow for FPGAs considers also options from the implementation phase of the VTA tensor accelerator overlay.} 
\label{fig2}
\end{figure}
}

One special contribution of TVM is, besides the large number of supported Frameworks and HW, to perform the so-called \dirquote{auto-tuning} of tensor operations in order to maximize performance~\cite{10.5555/3327144.3327258}. 
During auto-tuning, a number of known optimizations are performed on the Relay AST to improve the scheduling of the arithmetic operations. That way, the auto-tuning doesn't change the actual mathematical instructions -- like ALU or GEMM operations would -- of the program, but the execution order and memory accesses~\cite{10.5555/3327144.3327258}. TVM's current auto-tuning supports numerous optimizations across the lowering from DL code to an executable, categorized in high-level dataflow rewriting, memory reuse across threads, tensorized compute intrinsics and latency hiding. These are analytically presented in \cite{li2020deep}.

We propose that HW-SW interoperability be adopted to support the VTA auto-tuning in an agile way, inspired by the positive disruption of the Hardware Agile Manifesto~\cite{7436635}. As depicted in Fig.~\ref{fig2}, the processes of customizing the VTA 
overlay and optimizing the software in multiple IRs, are combined in a united auto-tuning task. With this approach we aim to extend TVM's current auto-tuning flow with HW design options for VTA in an automatic way. Most importantly, the intention of our approach is to select the most important HW options during the auto-tuning, particularly the ones correlated with different precision of VTA. In a nutshell, the ``agile'' concept is to provide a development environment where knowledge flows easily so the best solutions are reached as quickly as possible. Fig.~\ref{fig2} conceptualize the steps, where experience from DL development, DL software engineering and DL hardware engineering can be combined by spawning pipelines of tasks (i.e.~auto-tuning sprints) on multiple systems (i.e.~auto-tuning fleet) that share libraries, tools and devices in order to decrease auto-tuning time. This step will be further explained in subsection~\ref{collab}.

\subsection{Introducing the design space of $\tau$-VTA}

{
\setlength{\belowcaptionskip}{-1ex}
\begin{figure}[t]
\includegraphics[width=0.5\textwidth]{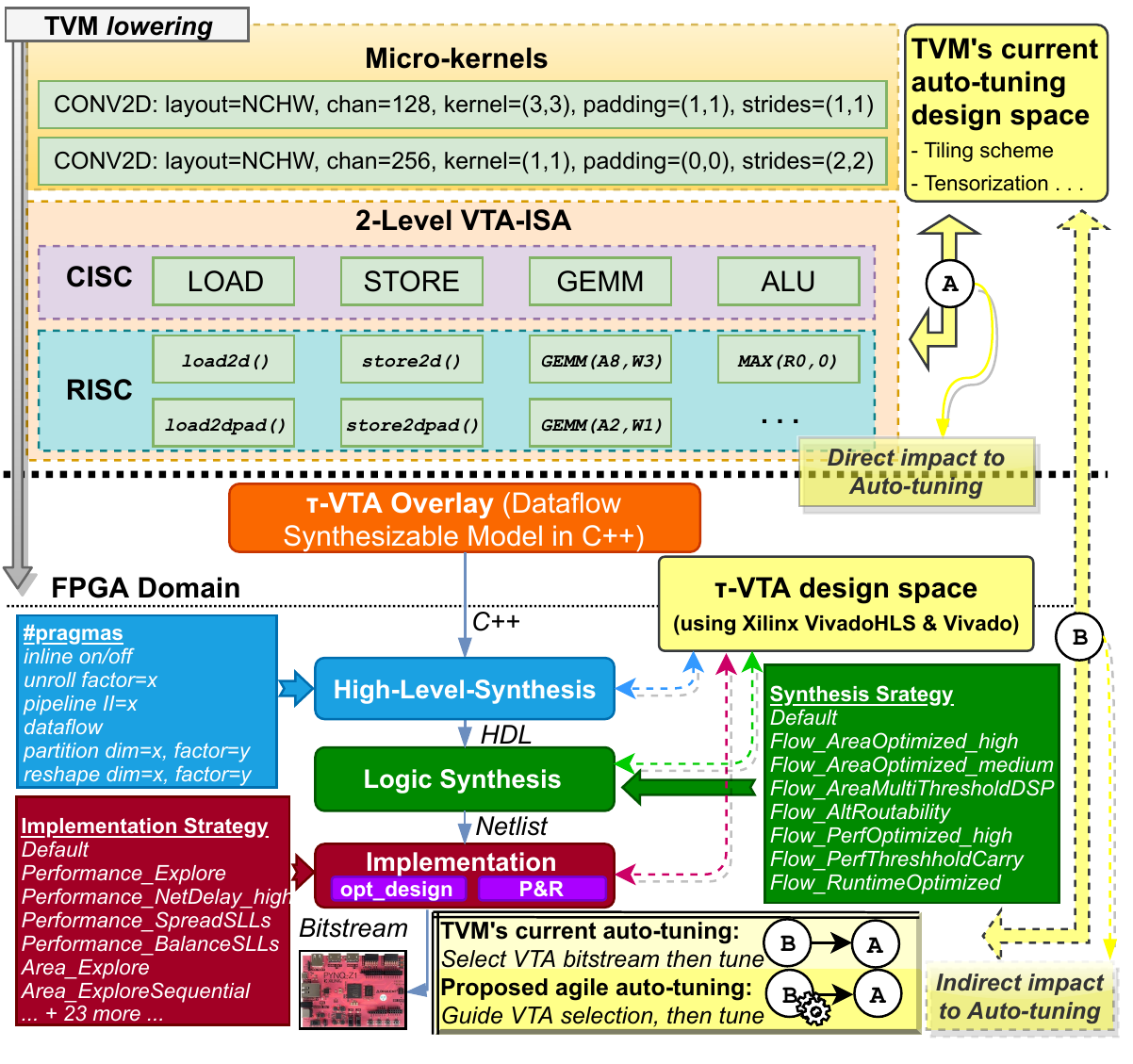}
\caption{The design space of $\tau$-VTA.}
\label{fig4}
\end{figure}
}

Figure~\ref{fig4} gives a high-level overview of the VTA hardware organization, as firstly presented in~\cite{8764458}. In addition, the figure highlights the main contribution of this work, i.e. the extension of TVM's auto-tuning design space with design options from HW expertise. VTA is composed of four modules: \textit{FETCH, LOAD, COMPUTE, and STORE}, which communicate over command queues and on-chip shared memories, implemented using FPGA block-RAMs (BRAMs). The \textit{FETCH} module dispatches task instructions after loading them from the DRAM. The \textit{LOAD} and \textit{STORE} units transfer tensor-tiles from DRAM into on-chip FPGA memories (BRAMs) and vice versa, respectively. The \textit{COMPUTE} unit performs operations on the register file. Specifically, the \textit{tensor ALU} performs element-wise tensor operations such as activation, normalization, and pooling tasks, while the  \textit{GEMM core} performs matrix multiplication over input and weight tensors. Common deep learning \textit{Micro-kernels}, such as 2D convolutions, are executed in \textit{GEMM core}~\cite{8764458}.

The VTA architecture is parameterizable, so that different shapes can be configured for tensor intrinsics, depending on the available resources. For example, the shape  of input, weight, and accumulator tensors that feed the GEMM tensor intrinsic unit directly affects the utilization of multipliers and the width of BRAMs ports. The data-types of the tensors are also parameterizable e.g. 8 bits or fewer. Programmability of the VTA is based on a two-level Instruction Set Architecture (ISA): i) a CISC-like task-ISA that explicitly orchestrates concurrent compute and memory tasks and ii) a RISC-like microcode-ISA that implements a wide variety of operators with single-cycle tensor-tensor operations. 

VTA is an overlay architecture programmed in synthesizable C++. Using the Xilinx Vivado HLS tool it is synthesized to a hardware description language (HDL) (either VHDL or Verilog) as a register-transfer-level (RTL) model. The downstream implementation stage, with the Xilinx Vivado tool, includes logic synthesis, place \& route, optimization (e.g. timing, area, energy) and the generation of a VTA bitstream for the FPGA device. 

All of these steps include many design choices that affect the trade-off between performance and resources utilization. In addition to the design options, the customization knobs of VTA define an additional hardware design space with 1000s of individual designs. VTA's developers explore which candidate to use in a sequence of steps. First, they use a simple FPGA resource model to prune unfeasible VTA parameterizations. After pruning, each candidate hardware design is compiled, placed, and routed. They select three tunable parameters, specifically, FPGA device, precision and batch size. Typically  their exploration  returns  a  handful  of promising candidates - ``the rest of the designs either yield low peak performance or fail placement, routing, or timing closure''~\cite{8764458}. For this final set of designs, they generate optimized software, using operator auto-tuning\cite{10.5555/3327144.3327258}, and use this software to obtain the workload's performance profile.

While VTA's optimization uses pruning instead of exhaustively exploring the design space to find the best candidate, the design space is limited to only three parameters. However, an overlay FPGA design can benefit from the knobs of design tools to deliver high performance. Such knobs may have an indirect impact on auto-tuning. For example, an optimal selection of the design options solely in the HLS step can lead to performance improvements of up to 29.030$\times$~\cite{DBLP:journals/corr/abs-1807-01340}. Figure~\ref{fig4} shows some important design options that are later discussed in Section~\ref{tvta}. We propose to find the impact of these parameters on VTA's performance by using a prediction technique and then guide the exploration based on the most important parameters. Hence, instead of pruning the design space with the hardware experience of designs that fail to meet the specs, we proactively offload the design space pruning to an algorithmic optimization problem (Section~\ref{tvta}). To establish a differentiation of the proposed VTA auto-tuning, from the current one in TVM stack, we use the term ``$\tau$-VTA auto-tuning'', accounting for transprecision.

{
\setlength{\belowcaptionskip}{-1ex}
\begin{figure}[t]
\includegraphics[width=0.5\textwidth]{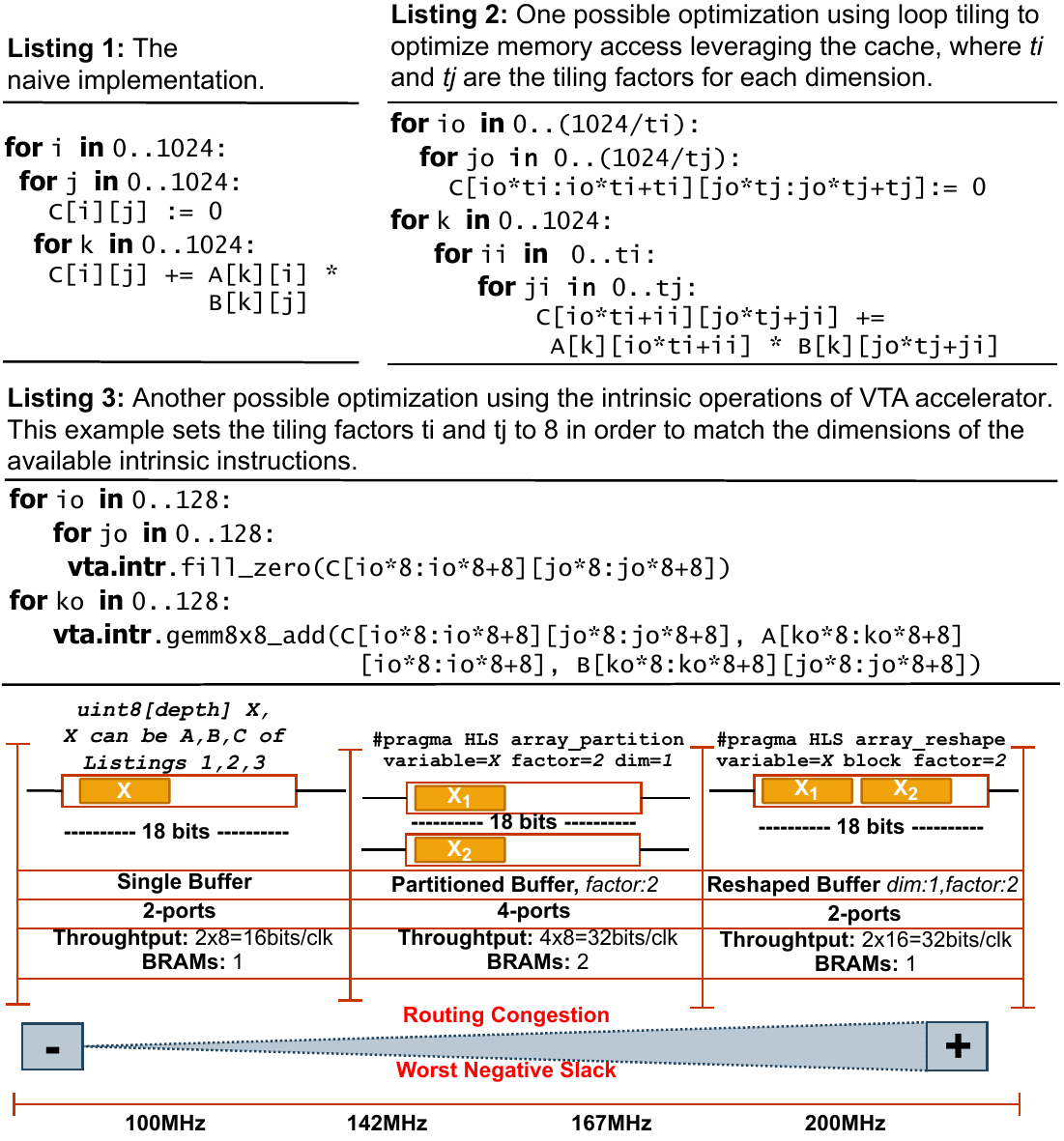}
\caption{Exemplary impact of features ``Partition'' and ``Reshape'' on memory throughput and BRAMs utilization for an $uint8$ array.}
\label{fig9}
\end{figure}
}

To highlight the difference of the current VTA auto-tuning by the ``$\tau$-VTA auto-tuning'', we present in code Listing~1 an example of a GEMM computation. The Listing~2 is expected to be faster on a target HW that uses caches, while the Listing~3 relies on intrinsic HW specific instructions. During TVM's auto-tuning phase, one of the optimizations is the selection of different tiling factors that accommodate for specific cache hierarchy or HW intrinsic dimensionality on the HW device (VTA). This selection can be either completely random or algorithmic-driven as shown in~\cite{10.5555/3327144.3327258}, where XGBoost~\cite{Chen:2016:XST:2939672.2939785} and TreeGRU~\cite{Tai2015ImprovedSR} algorithms have shown to find better code-refactoring candidates in shorter time. However this analysis neglects some important features of the reconfigurable HW. For example, if the data-type of the computation is $uint8$, then during the design of VTA we can use the Vivado HLS directives \textit{\#pragma HLS ARRAY\_PARTITION} and \textit{\#pragma HLS ARRAY\_RESHAPE} to increase the memory throughput and BRAMs utilization, as shown in Fig.~\ref{fig9}, assuming 18Kbits true dual-port Xilinx BRAMs. The case of $uint4$ will double the throughput and so on. However, the aggressive use of those directives is known to increase the routing congestion and the difficulty to meet the target frequency. Such trade-offs are known to the HW engineering community, but not in the SW counterpart. The ``$\tau$-VTA auto-tuning'' bridges this gap by algorithmically selecting the most important VTA HW parameters, for different precision settings, before initiating the TVM's current VTA auto-tuning.


\subsection{Collaborative Exploration Pipeline}\label{collab}

An end-to-end DNN framework, such as TVM, combines expertise from many levels of the computing stack, i.e. DNN front-end languages, compilers, IRs, scheduling, HW generation, etc. With it being a community-driven framework, it is expected that updates are introduced in a dynamic, non-deterministic time-plan. Automating the integration of code changes from different contributors is a key element for maintaining project stability. In this work we explore the optimal VTA designs for a given precision, based on features of the hardware design space. We propose the use of tools that allow us to enforce the concept of Continues Integration (CI) during this exploration. 

Figure~\ref{fig:docker-flow} shows the overview of this CI exploration pipeline. The proposed CI infrastructure consists of a cluster of machines, named auto-tuning fleet, on which we deploy our exploration jobs. The key technologies we use are Jenkins, Docker and Docker Swarm. \textbf{Jenkins} is the tool we use to define, run and manage the CI jobs. Jenkins supports the ``Master+Agent'' mode, where the Master is in charge of orchestration and acts as the user end-point, and the Agents perform the actual work, i.e. exploration. These agents can be distributed across many servers. Thanks to the Jenkins plug-in that provides integration with Docker, the agents can be even spawned on-demand as containers. \textbf{Docker} is the supporting technology for the whole infrastructure. Every component runs as a Docker container, even the Jenkins Master itself. This establishes portability and scalability for the exploration phase. \textbf{Docker Swarm} allows a seamless deployment of containers on any of the cluster's machines in a transparent way. It is responsible for load-balancing and tracking of the status of all running containers, across all of the machines that have joined the Swarm.

The advantage of the proposed collaborative exploration pipeline is that, whenever a community member commits updates of the VTA overlay to the TVM repository (e.g. new GEMM unit, new precision in accumulator etc.), the exploration of the optimal configurations of the VTA design space is triggered automatically. As soon as an optimal VTA bitstream is explored, auto-tuning is triggered as a subsequent step, named auto-tuning sprint (inspired by the agile concept). The best VTA bitstreams and auto-tuning configurations are stored in the local repository and they can be uploaded back to the community's repository with a merge request.

{\setlength{\belowcaptionskip}{-3ex}
\begin{figure}[htbp]
\includegraphics[width=0.5\textwidth]{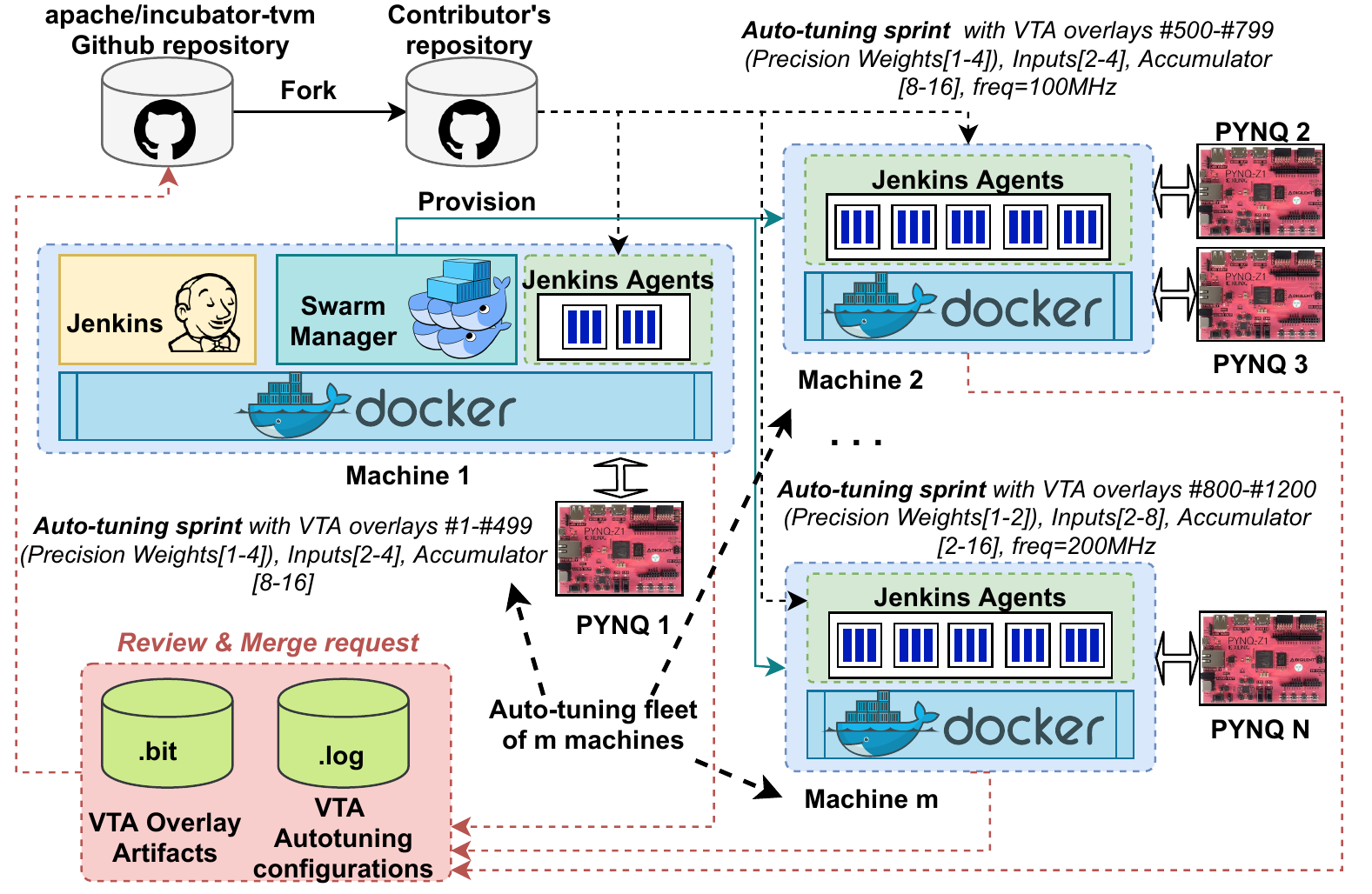}
\caption{The exploration pipeline using Docker Swarm and Jenkins.}
\label{fig:docker-flow}
\end{figure}
}

%% file: optimize.tex
\section{$\tau$-VTA auto-tuning}\label{tvta}
\subsection{$\tau$-VTA Features}

\subsubsection{Feature Selection}\label{feature_sel} 
To enable a guided VTA design space exploration, by annotating knowledge from HW experience, we build a dataset of HLS and implementation results, consisting of samples across individual designs. We run each design through the complete C-to-bitstream flow for various design options. These options are used in addition to the ones of current TVM's auto-tuning~\cite{8764458}, i.e. fpga device, precision and batch size. The design options are the features of our formulation for predicting an optimization target. In this analysis we explore two main targets, performance, in terms of Giga Operations per second (GOPs) and resources utilization, in terms of absolute number of utilized FPGA resources. For the latter target we explore every resource individually, i.e. Block RAMs (BRAM)s, Flip Flops (FFs), Digital Signal Processors (DSPs) and Look-Up-Tables (LUTs). Feature selection is a very important step for our flow, so we decide not only to include the design options commonly used by the FPGA community, based on our experience, but also the ones used in research around design space exploration with EDA tools~\cite{DBLP:journals/corr/abs-1807-01340, 8457644}. Table~\ref{tab:features} lists the features we've selected in this research. Similarly, we extract implementation results, known as the targets in our optimization problem, from the implementation reports. After extraction, our dataset contains features and targets for each design sample and can be used to develop prediction models that map from features to targets.

\begin{table}[htbp]
	\caption{Features used in the architecture search}
	\begin{center}
	\footnotesize
	\scalebox{0.9}{
		\begin{tabular}{L{1.3cm}|L{5.1cm}|L{2.1cm}}
			 \textbf{Feature} & \centering\textbf{Description} &  \textbf{Search range}\\
			\hline 
			\hline
			HLS\_freq & The frequency Vivado HLS is aiming for.&   [100:20:500]MHz \\ \hline
			Inline & Flatten RTL hierarchy by function inlining.&  Enabled/Disabled\\ \hline
			Unroll & Unroll a loop with a given factor.& [1,2,4,8,complete] \\ \hline
			Pipeline & Pipeline(II=1) design with registers.&  Enabled/Disabled \\ \hline
			Dataflow & Use task-level pipelining. & Enabled/Disabled \\ \hline
			Partition & Distribute the memory contents across multiple BRAMs. & cyclic/block dim=[1,2] factor=[1:32]\\ \hline
			Reshape & First Partition, then rejoin BRAMs to decrease utilization.& cyclic/block dim=[1,2] factor=[1:32]\\ \hline 
			Impl\_freq & The frequency during place and route.& [100:20:500]MHz \\ \hline
			Syn\_strategy & The Vivado synthesis strategy (heuristic).& All available (8) \\ \hline
			Imp\_strategy & The Vivado implementation strategy (heur.).& All available (32) \\ \hline
			\hline
		\end{tabular}}
		\label{tab:features}
	\end{center}
\end{table}

\subsubsection{Irrelevant Features Elimination}\label{irrelevant}
Some of the design options are correlated. As such, combinations of features that are known a priori to exert little influence on the targets should be eliminated to reduce the dimensionality of the data. Having fewer features leads to simpler models, which require shorter training time, and reduces the chance of over-fitting. 
Feature elimination can benefit from HW knowledge. 
However, even if some feature choices are valid from a HW knowledge perspective, they may lead to unfeasible designs. For example, revisiting the example of Figure~\ref{fig9}, the large factors of partitioning/reshaping can lead to more congested routing, which in combination with the frequency requirements, may result in unsuccessful timing closure. We eliminate irrelevant features by formulating them as constraints in the $\tau$-VTA optimization problem.

\subsection{Problem Formulation}

\subsubsection{Prediction Model for Feature Importance}\label{model}

We train a classification model to predict the impact of features on the optimization goal of increasing performance while respecting constraints in resources utilization. For our study, we have a set of $n$ training samples $\left\{\mathbf{x}_{i}, \mathbf{y}_{i}\right\}_{i=1}^{n}$, where $\mathbf{x}_{i}=\left[x_{i}^{1}, x_{i}^{2}, \ldots, x_{i}^{p}\right]^{\top} \in \mathbb{R}^{p}$ is the input vector of feature values for the $i$th sample, and $\mathbf{y}_{i}=\left[y_{i}^{1}, y_{i}^{2}, \ldots, y_{i}^{q}\right]^{\top} \in \mathbb{R}^{q}$ is the corresponding vector of target values. $p$ denotes the number of input features (e.g., HLS\_freq, Unroll, precision etc.), and $q$ denotes the number of output targets (i.e., actual GOP, LUT, FF, DSP, and BRAM counts after the Vivado-implementation). In addition we define $\mathbf{X}=\left[\mathbf{x}_{1}, \dots, \mathbf{x}_{\mathbf{n}}\right]^{\top}$ to denote feature values for all samples and $\mathbf{y}^{\mathbf{k}}=\left[y_{1}^{k}, y_{2}^{k}, \ldots, y_{n}^{k}\right]^{\top}$ to denote values of the target $k$ for all samples.

Each learning task corresponds to one target prediction. We train a separate model $f_{k}$ for each target $k$, resulting in a set of mapping functions $\left\{f_{k}: \mathbb{R}^{p} \rightarrow \mathbb{R}\right\}_{k=1}^{q}$. We select the gradient tree boosting algorithm to build our prediction model. Specifically, we model the target as the sum of regression trees, each of which maps the features to a score for the target. Target estimation is determined by accumulating scores across all trees. By implementing gradient descent, gradient tree boosting optimizes the loss over the space of regression trees by repeatedly selecting the tree that points in the negative gradient direction. Since we eliminate irrelevant features (Section~\ref{irrelevant}), many design options result in unfeasible implementations (e.g. timing violation, resource limitations etc.), which result in very sparse datasets. We propose the use of XGBoost~\cite{Chen:2016:XST:2939672.2939785}, a gradient tree boosting algorithm for sparse datasets that employs a sparsity-aware approximate split finding technique. 

The benefit of using XGBoost is that, after the boosted trees are constructed, it is relatively straightforward to retrieve importance scores for each feature and use them for our mapping functions $\left\{f_{k}: \mathbb{R}^{p} \rightarrow \mathbb{R}\right\}_{k=1}^{q}$. Importance provides a score that indicates how valuable each feature was in the construction of the boosted decision trees within the model. The more a feature is used to make key decisions in decision trees, the higher its relative importance. This importance is calculated explicitly for each feature in the dataset. For a single decision tree, importance is calculated through the amount that each feature-split-point improves the performance measure by, weighted by the number of observations the node of the tree is responsible for. The feature importances are then averaged across all of the decision trees within the model to obtain the feature importance space $\mathcal{F}_{i}$.

\subsubsection{$\tau$-VTA Auto-tuning Algorithm}\label{algo}
For a given tensor operator specification, e.g. Listing~1, there are multiple possible low-level program implementations, each with different choices of loop order, tiling size, and other options, e.g. Listing~2 and 3. The problem of $\tau$-VTA auto-tuning is to find which logically equivalent programs exhibit fastest execution for a VTA overlay of selected precision. Towards this goal, we extended TVM's auto-tuning algorithm with an extra outer loop that iterates over a number of possible optimal VTA overlays. This pool is populated with VTA bitstreams, by giving priority to the configurations of the design features with higher importance per precision target. Algorithm 1 describes this algorithmic behavior. The lines 4-to-9 correspond to TVM's auto-tuning algorithm~\cite{8764458} and describe the steps needed to obtain an optimal configuration from a software transformation space $\mathcal{S}_{e}$, on a target hardware device. Specifically, lines 5-to-8 account for combining quality and diversity when selecting b candidates for hardware evaluation and are analytically explained in~\cite{8764458}. The iteration of these steps over an outer loop (line 1) establishes the search for an optimal VTA overlay for a given precision $p$. Thus, the difference of Algorithm~1 with the auto-tuning algorithm in~\cite{8764458}, is that instead of performing auto-tuning only with one VTA overlay, the Algorithm 1 proposes the auto-tuning in multiple overlays, in parallel. The output is not only an optimal configuration from a software transformation space $\mathcal{S}_{e}$, but also a VTA overlay $o^{*}_{p}$ for a precision requirement $p$. The overlays with the highest probability of being optimal are prioritized in this exploration using the feature importance metric, which is described in previous subsection~\ref{model}.

\begin{algorithm}[t]
\small
\DontPrintSemicolon
 
  \KwInput{VTA precision $p$}
  \KwInput{Feature importance space $\mathcal{F}_{i}$ (discussed in~\ref{model})}
  \KwInput{Software transformation space $\mathcal{S}_{e}$}
  \KwOutput{Selected VTA overlay $o^{*}_{p}$ for precision $p$}
  \KwOutput{Selected schedule configuration $s^{*}_{p}$ for precision $p$}
  \KwData{Database $\mathcal{D}_{o}$ of run time statistics for VTA overlay $o$}
 
 \While{k\_trials\_overlays $<$ max\_k\_trials\_overlays}   
 { 
 \tcc{Auto-tuning sprints on a fleet of $m$}
 ${M_p} \leftarrow$ run $m$-parallel jobs of VTA bitstream generation to collect candidates in $o^{*}_{p}$, with higher probability for the most important features of $\mathcal{F}_{i}$ for precision $p$
 
 $\mathcal{D}_{o} \leftarrow \emptyset$
 
 \While{n\_trials $<$ max\_n\_trials} 
  {
   \tcc{Pick the next promising batch}
  
   $Q \leftarrow$ run parallel simulated annealing to collect candidates in $\mathcal{S}_{e}$ for the VTA overlay ${M_p}$
  
  

   $S \leftarrow$ select $(1-\epsilon) b$-subset from $Q$ using Random Search or TreeGRU~\cite{Tai2015ImprovedSR} optimizer (as in current VTA auto-tuning)
   
   $S \leftarrow S \cup\{\text { Randomly sample } \epsilon \text { candidates. }\}$

   Run $b$ measurements on VTA configured with ${M_p}$
  
  
  $\mathcal{D}_{o} \leftarrow $ exec. time \tcp*{Update best trials}
  
  n\_trials $\leftarrow$ n\_trials $+ b$

  }
  $s^{*}_{p} \leftarrow$ history best schedule configuration for precision $p$
  
  
  k\_trials\_overlays $\leftarrow$ k\_trials\_overlays $+ m$
  }
  
 $o^{*}_{p} \leftarrow$ history best VTA overlay for precision $p$
 
\caption{\small{Learning to Optimize TP Tensor Programs}}
\end{algorithm}



%% file: expt.tex

\section{Evaluation Results}\label{results}
\subsection{Feature Importance}\label{results_feature}

{\setlength{\belowcaptionskip}{-0ex}
	\begin{figure*}[htbp]
		\includegraphics[width=1.0\textwidth]{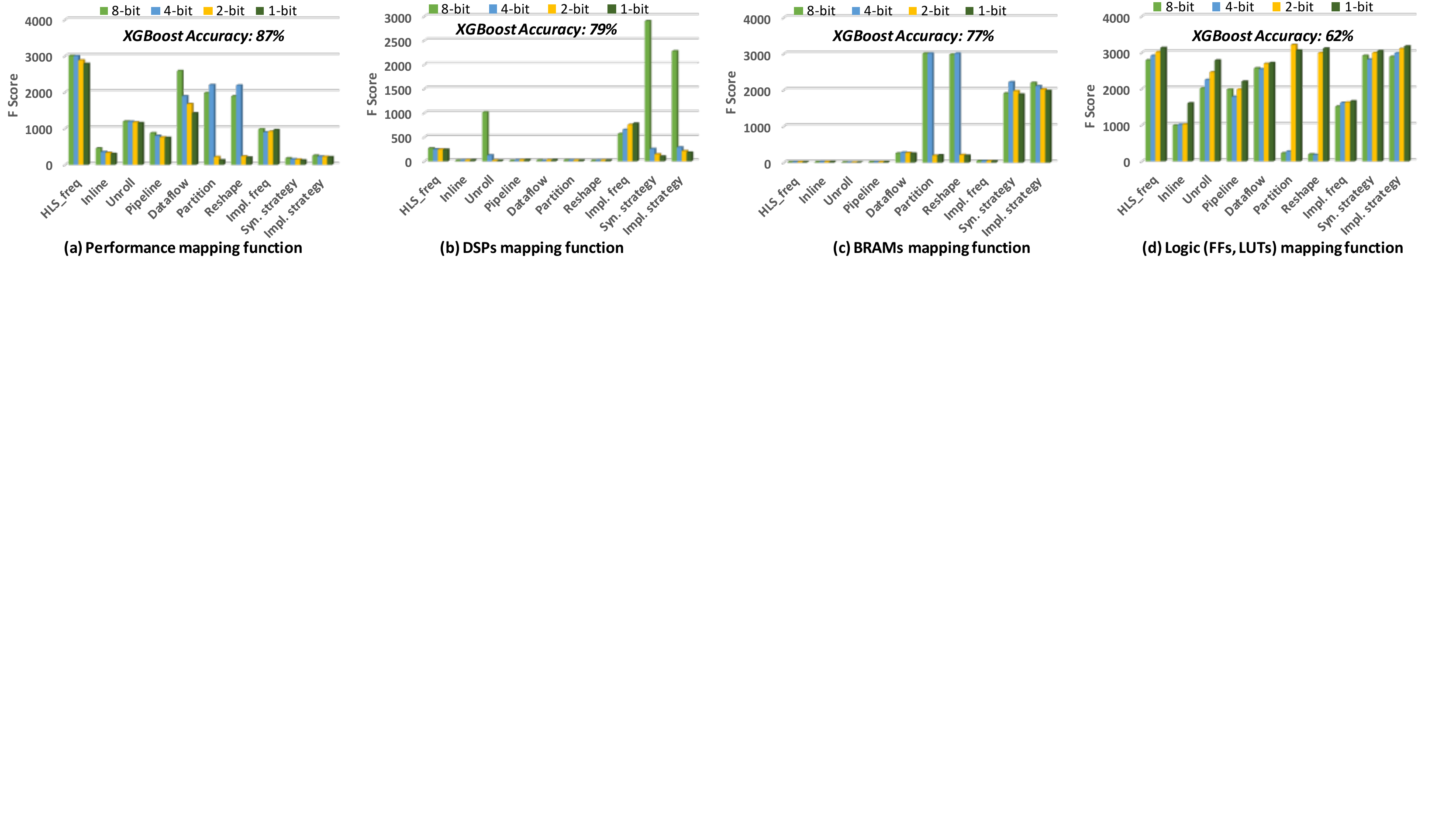}
		\vspace{-210pt}
		\caption{Features' impact on the HW design space of $\tau$-VTA, using an XGBoost model. Different features express the diverse impact on performance and resources with respect to precision specification.}
		\label{fig:fig6}
	\end{figure*}
}

We evaluate the feature importance prediction model with a design space defined by the values listed in Table~\ref{tab:features}. In addition, we use a transprecision vector of [$1,2,4,8$] bits. In this analysis, transprecision is defined as the flexibility to change precision on-demand by the application. This is a typical requirement in the area of NN architecture exploration ~\cite{2019arXiv190106261S}. The initial design space includes 797,442,048 evaluations (\textit{design options listed in Table~\ref{tab:features}} $\times$ [$1,2,4,8$] bits). After the elimination of irrelevant features (Section~\ref{irrelevant}), the design space contains 2,440 designs. After training the model described in Section~\ref{model} with a subset of the design space, we build the prediction model to find the importance of design features for performance and resources. Then, we use this model to select the VTA configurations that will be implemented for a given input vector (precision, freq, etc.).

We implement and train the prediction model in Python leveraging the XGBoost library~\cite{Chen:2016:XST:2939672.2939785}. All designs in the dataset are synthesized and implemented with Xilinx Vivado 2018.3 (VTA-supported) targeting Pynq-Z1. Experiments are performed on an Intel Xeon E5-2630 processor running at 2.6GHz. The models is able to complete the prediction tasks within seconds, compared to the minutes or hours that each implementation typically incurs. We measured 35 minutes for a single end-to-end implementation with default settings for VTA, Vivado HLS and Vivado. Some design options can lead to implementations that last up to eight hours. The average prediction time for all mapping functions is 2.7 seconds.

{\setlength{\belowcaptionskip}{-2ex}
	\begin{figure*}[htbp]
		\includegraphics[width=1.0\textwidth]{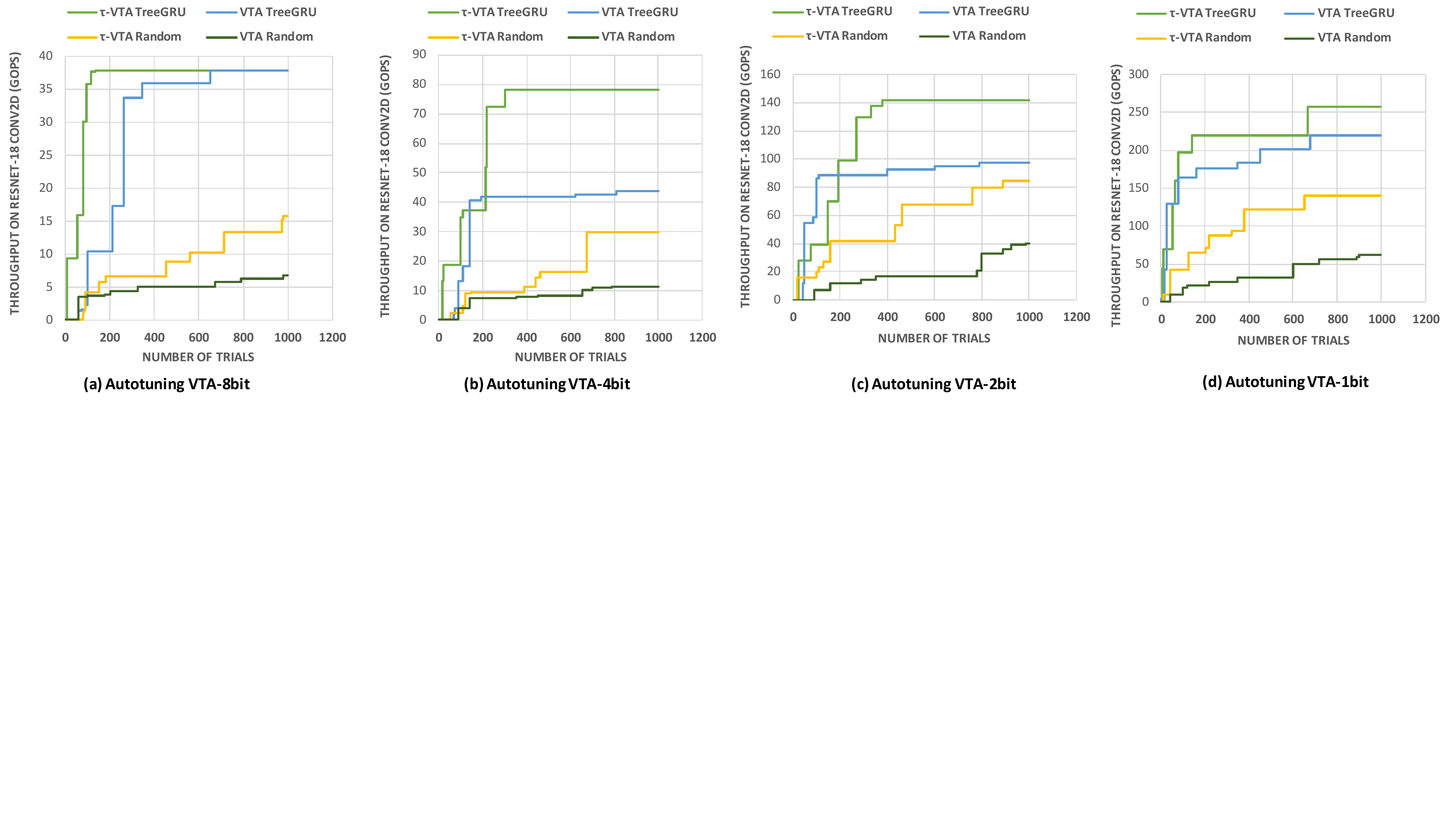}
		\vspace{-160pt}
		\caption{Schedule exploration with a na\"ive random search and TreeGRU algorithm for a single connvolution layer on PYNQ-Z1. Original VTA and $\tau$-VTA are compared. Design candidates with (2,16)x(16,16)and (8,8)x(8,8) GEMM intrinsic at a)W8A8 a)W4A4 a)W2A2 d)W1A1 are considered. The layer is conv2d: IC=256, OC=256, H=W=14, KW=KH=3, stride=(1,1),padding=(0,0).}
		\label{fig:fig7}
	\end{figure*}
}

For importance classification, we compute the percentage of incorrectly classified samples out of the total number of samples. We randomly select 20\% of our data as the testing set and perform a cross-validation by random permutation over 10 iterations on the remaining training/validation set. In each iteration, we randomly select 75\% of the training/validation set for training and 25\% for validation. The validation set is used for parameter tuning to locate better VTA models for implementation. Figure~\ref{fig:fig6} depicts the feature importance on the HW design space of $\tau$-VTA, for the objective targets a)~Performance(GOPs), b)~DSPs, c)~BRAMs and d)~Logic resources (FFs, LUTs). For every objective we plot the importance score $F$ for every feature. The precision feature is plotted as a grouped option. Please note that the plotting is orthogonal to all features and any combination can be used. Since we focus on transprecision overlays in this study, we select an analysis that expresses the precision as a grouping option over the others. 

Figure~\ref{fig:fig6} also shows the prediction accuracy. DSPs and BRAMs are generally easier to estimate than logic resources because operations with LUTs and FFs experience more complicated transformations than coarse-grain elements like DSP and BRAMs. In general, XGBoost has an accuracy of 87\% for predicting the importance of performance and 72\%, 77\% and 68\% for DSPs, BRAMs, and logic, respectively.

\subsection{$\tau$-VTA Auto-tuning}
After the feature importance analysis, we guide the selection of the final set of designs and generate optimized software using operator auto-tuning~\cite{10.5555/3327144.3327258}. Component evaluations were based on convolution workloads in ResNet-18 for ImageNet classification. The accuracy evaluation of ResNet-18 is orthogonal to this analysis since we don't aim to optimize the accuracy, but to optimize the performance of the convolutions given a predefined precision. Hence, the precision choice is used as input for the optimization of the $\tau$-VTA overlay and not for its accuracy, which is a separate step. Figure~\ref{fig:fig7} compares the performance of convolutions when the VTA designs are selected by using the current state-of-the-art approach (VTA) or by using the feature importance metric ($\tau$-VTA). We select two auto-tuning methods from~\cite{10.5555/3327144.3327258} which are implemented in TVM, random search and TreeGRU~\cite{Tai2015ImprovedSR}. For the case of $\tau$-VTA we select the trained prediction model of Section~\ref{results_feature} and we get the predicted VTA designs that are proposed by the model to operate better, for an input feature vector of a precision of [$1,2,4,8$] bits. We set the maximum number of iterations $n=1000$. Our proposed $\tau$-VTA designs are able to guide the auto-tuning to higher performance for the transprecision vector of [8,4,2,1] bits by 1$\times$, 1.7$\times$, 1.4$\times$ and 1.2$\times$ for TreeGRU and 2.3$\times$, 2.5$\times$, 2.1$\times$ and 2.3$\times$ for random search. In addition, the $\tau$-VTA auto-tuning converges faster by 1.5$\times$, 5$\times$, 5.2$\times$ and 7.1$\times$ for TreeGRU and 6.6$\times$, 2.2$\times$, 6.2$\times$ and 8.1$\times$ for random search, respectively.


%% file: related.tex
\vspace{-1pt}
\section{Related work}
\vspace{-1pt}
To accelerate DL models on diverse DL HW, it is important to map the computation to DL HW efficiently. On general-purpose HW, efficient computation of DL models relies on the highly optimized linear algebra libraries such as Basic Linear Algebra Subprograms (BLAS) libraries (e.g., MKL and cuBLAS). In addition, the HW vendors have released specially optimized libraries tailored for DL computations (e.g., cuDNN and MKL-DNN). More advanced tools, like TensorRT~\cite{tensort}, support graph optimization and low-bit quantization with large collection of highly optimized GPU kernels. Due to enormous search space for parameter tuning in hardware-specific optimizations, an emerging set of techniques, namely auto-tuning, are necessary to determine the optimal parameter settings. Apart from TVM's auto-tuning, which we used in our analysis, and was used with FPGAs~\cite{8764458}, as well as GPUs and embedded devices~\cite{10.5555/3327144.3327258}, numerous DL works employ optimizers for parameter tuning. Tensor Comprehensions~\cite{DBLP:journals/corr/abs-1802-04730} firstly uses black-box optimization to select parameters of thread blocks and secondly polyhedral optimization to generate internal loops. This task, falls within the category of hyper-parameter optimization for optimal code generation for different HW back-ends. An extensive review of such optimizers, alongside the DNN compilation frameworks to apply these optimizers, is presented in~\cite{li2020deep}. Similar to our approach, the formulation of a prediction model that quantifies the impact of a hardware design choice, has been approached several times in literature already~\cite{8847448}. The general category of such research direction falls within the design space exploration of HLS tools for an FPGA model. However, to the best of our knowledge, the only background work for auto-tuning with TVM compiler stack, is~\cite{10.5555/3327144.3327258}. As a future work we aim to study how our prediction model depends on a particular DNN topology. 



%% file: conclusion.tex
\section{Conclusion}\label{conclusion}
This research explores the possibility of automatically guiding the auto-tuning of an overlay architecture, for different transprecision settings, by leveraging knowledge from hardware experience. By adopting the concept of agile development we built a pipeline of engineering tasks that support the auto-tuning process. Instead of eliminating the overlay hardware design space with pruning techniques, we propose a technique that builds a prediction model to quantify the impact of a hardware design choice towards an optimization goal. We show that the features with the highest impact differ for different precisions. By using the most important features in order to generate an overlay we manage to perform auto-tuning that succeeds in higher performance of up to 2.5$\times$ and faster convergence of up to 8.1$\times$.